%% file: zpol.tex
\documentclass[twocolumn,letterpaper,aps,prl,superscriptaddress,showpacs,floatfix,preprintnumbers,longbibliography]{revtex4-1}

\usepackage{graphicx}   
\usepackage{xspace}     
\usepackage{url}
\usepackage[mathlines]{lineno}
\usepackage{bm}
\usepackage{amsmath}
\usepackage{xcolor}
\usepackage[dvipdfm]{hyperref}
\hypersetup{
	colorlinks = true,
	citecolor= blue,
	linkbordercolor = cyan
}
\newcommand{\pt}{\mbox{$p_T$}\xspace}

\newcommand{\sqsn}{\mbox{$\sqrt{s_{_{NN}}}$}\xspace}

\newcommand{\lam}{\mbox{$\Lambda$}\xspace}
\newcommand{\alam}{\mbox{$\bar{\Lambda}$}\xspace}
\newcommand{\costhe}{\mbox{$\langle\cos\theta_p^{\ast}\rangle$}\xspace}
\newcommand{\mean}[1]{\langle #1 \rangle}
\usepackage{color}
\definecolor{orange}{cmyk}{0.,0.353,1.,0.}    

\begin{document}
%

\title{Polarization of $\Lambda$ ($\bar{\Lambda}$) hyperons along the beam direction in
  Au+Au collisions \\ at \sqsn = 200 GeV}

\input{authorlist_05282019}

\date{\today}

\begin{abstract} 
The $\Lambda$ ($\bar{\Lambda}$) hyperon polarization along the beam direction 
has been measured for the first time in Au+Au collisions at
$\sqrt{s_{_{NN}}}$ = 200 GeV. The polarization dependence on the
hyperons' emission angle relative to the second-order event plane
exhibits a sine modulation, indicating a quadrupole pattern of the
vorticity component along the beam direction.
The polarization is found to increase in more
peripheral collisions, and shows no strong transverse momentum ($p_T$)
dependence at $p_T>1$ GeV/$c$.  The magnitude of the signal is about
five times smaller than those predicted by hydrodynamic and 
multiphase transport models; the observed phase of the emission angle
dependence is also opposite to these model predictions.  In contrast,
blast-wave model calculations reproduce the modulation phase measured
in the data and capture the centrality and transverse momentum
dependence of the signal once the model is required to reproduce
the azimuthal dependence of the Gaussian source radii measured via
the Hanbury-Brown and Twiss intensity interferometry technique.
\end{abstract}

\maketitle


\setlength\linenumbersep{0.10cm}

The properties of deconfined partonic matter, the quark-gluon plasma, have been
explored in heavy-ion collisions at the Relativistic Heavy Ion
Collider (RHIC)~\cite{wh_star,wh_phenix,wh_phobos,wh_brahms} and the Large
Hadron Collider~\cite{Raa_alice,jet_cms,dijet_atlas}.  The matter
created in non-central heavy-ion collisions should exhibit rotational
motion in order to conserve the initial angular momentum carried by the two 
colliding nuclei. The direction of the angular momentum is perpendicular to the
reaction plane, as defined by incoming beam and the impact parameter
vector.  It was predicted~\cite{Liang_2005,Voloshin:2004ha} that such
a spinning motion of the matter
would lead to a net spin polarization of particles produced in the
collisions due to spin-orbit coupling. Hyperons are natural candidates
to explore this phenomenon since in the parity violating weak decays of
the hyperons the momentum vector of the decay baryon is highly
correlated with the hyperon spin.
In such decays the angular distribution of the daughter baryons 
is given by:
\begin{eqnarray}
\frac{dN}{d\cos\theta^{\ast}} \propto 1 
+ \alpha_{H}P_{H}\cos\theta^{\ast}, \label{eq:hyperon_decay}
\end{eqnarray}
where $\alpha_H$ is the hyperon decay parameter, $P_H$ is the hyperon
polarization, and $\theta^{\ast}$ is the angle between the
polarization vector and the direction of the daughter baryon momentum
in the hyperon rest frame.  

The Solenoidal Tracker at RHIC (STAR) Collaboration has observed 
positive polarizations of \lam
hyperons along the orbital angular momentum in Au+Au collisions for
collision energies of \sqsn = 7.7 -- 200
GeV~\cite{polBES,Adam:2018ivw}.  This polarization is evidence for the
creation of the most vortical fluid ever observed, with vorticities of
the order of $\omega\sim10^{22}~s^{-1}$. These results open new
opportunities for a better understanding of the dynamics and
properties of the matter created in heavy-ion collisions.

The spin polarization of hyperons along the orbital angular momentum
of the entire system is referred to as the {\em global} polarization,
meaning a net spin alignment along a globally defined direction.
However, the vorticity and, consequently, the particle polarization
may vary for different regions of the fluid due to anisotropic flow,
energy deposits from jet quenching, density fluctuations, etc. The
detailed structure of the vorticity fields may be complicated and the
resulting particle polarization can depend on the particle transverse
momentum and the azimuthal angle relative to the reaction plane, or
even exhibit toroidal structures
~\cite{Betz:2007kg,Voloshin:2017kqp,Becattini:2017gcx,Pang:2016igs}.

\begin{figure}[hbt]
\begin{center}
\includegraphics[width=0.65\linewidth]{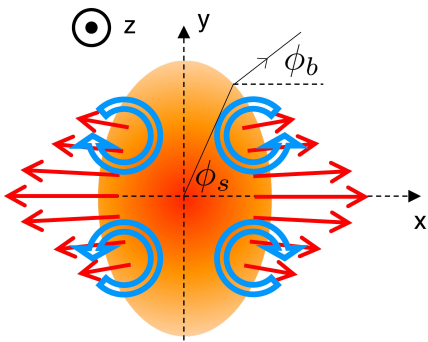}
\caption{\label{fig:cartoon}(Color online) A sketch illustrating the
  system created in a non-central heavy-ion collision viewed in the
  transverse plane (x-y), showing stronger in-plane expansion (solid
  arrows) and expected vorticities (open arrows). 
  In this figure the colliding beams are oriented along the z-axis and the x-z plane defines the reaction plane.
  See text for explanations of $\phi_s$ and $\phi_b$. 
}
\end{center}
\end{figure}

Anisotropic flow, characterized by the Fourier coefficients of the
particle azimuthal distribution in the transverse plane, has been
extensively studied in heavy-ion collisions and was found to be well
described by hydrodynamic
calculations~\cite{Voloshin:2008dg,Heinz:2013th}.  Nontrivial velocity
fields describing transverse anisotropic flow should lead to a
vorticity component along the beam direction dependent on the
azimuthal angle relative to the reaction
plane~\cite{Voloshin:2017kqp,Becattini:2017gcx}.  The observation of
the large second-order coefficients, a.k.a. elliptic flow, 
in mid-central collisions indicates significantly
stronger expansion in the reaction plane direction compared to that
out-of-plane, which might lead to a quadrupole structure in the
$z$-component of vorticity as illustrated in Fig.~\ref{fig:cartoon}.
Experimental measurements of such a component are the main goal of this
analysis.

The beam direction component of the polarization arising from vorticity 
due to elliptic flow is expected to be more sensitive to later times 
from flow development in the system
evolution~\cite{TeaneyYan_2011}, unlike the global polarization that originates mostly from
the initial velocity fields. It might also have different sensitivity
to the relaxation time needed for the conversion of the vorticity into
particle polarization. Therefore, it is of great interest to study the
polarization along the beam direction for further understanding of the
role of the vorticity in heavy-ion collisions and possibly to answer
these questions. In this Letter, we report the beam direction component
of polarization for \lam and \alam hyperons in Au+Au collisions at
\sqsn = 200~GeV. The results are presented as functions of the
collision centrality and hyperons' transverse momentum ($p_T$).

The dataset for this analysis was collected in 2014 by 
the STAR detector during the period 
of Au+Au collisions at \sqsn = 200~GeV.
Charged-particle tracks
were measured in the time projection chamber (TPC)~\cite{tpc}, which
covers the full azimuth and a pseudorapidity range of $-1<\eta<1$.
The collision vertices were reconstructed using the measured
charged-particle tracks.  Events were selected to have the collision
vertex position within 6~cm of the center of the TPC in the beam
direction and within 2~cm in the radial direction with respect to the
beam center. In addition, the difference between the vertex positions
along the beam direction determined by the TPC and the vertex position
detectors (VPD)~\cite{vpd} located at forward and backward rapidities
($4.24<|\eta|<5.1$) was required to be less than 3~cm to suppress pileup 
events. These selection criteria yielded about one billion minimum 
bias events, where the minimum bias trigger required hits of both 
VPDs and the zero-degree calorimeters~\cite{zdc} located at $|\eta|>6.3$.

The collision centrality was determined from the measured
multiplicity of charged particles within $|\eta|<0.5$ and a
Monte-Carlo Glauber simulation~\cite{BESv2}.  The second-order event
plane ($\Psi_2$) as an experimental estimate of the reaction plane was
determined by the charged-particle tracks within the transverse
momentum range of $0.15<\pt<2$ GeV/$c$ and $0.1<|\eta|<1$:
\begin{eqnarray}
\Psi_2^{\rm obs} = 
\tan^{-1}\left(\sum_i w_i\sin(2\phi_i)/\sum_i w_i\cos(2\phi_i)\right), 
\label{eq:psi2}
\end{eqnarray}
where $\phi_i$ and $w_i$ are the azimuthal angle and \pt of the $i^{\rm th}$ particle in the event.
The resolution of the measured plane $\Psi_2^{\rm obs}$ defined
as ${\rm Res}(\Psi_2)=\langle\cos2(\Psi_2^{\rm obs}-\Psi_2)\rangle$ was
estimated with the two-subevent method~\cite{TwoSub}, where the two
subevents were taken from $0.1<|\eta|<1$. 
In mid-central collisions the event plane resolution peaks at $\sim$0.76. 

Charged-particle tracks reconstructed with the TPC were selected to
have good quality by requiring the following conditions. The number of hit points used in the
track reconstruction was required to be larger than 15. The ratio of
the number of hit points used to the maximum possible number of TPC
space points for that trajectory was required to be larger than
0.52. Tracks within $0.15<\pt<10$~GeV/$c$ and $|\eta|<1$ that passed
through the track selections above were used to reconstruct \lam
hyperons.  In order to reconstruct \lam and \alam, the decay channels
of $\lam\rightarrow p+\pi^{-}$ and $\alam\rightarrow \bar{p}+\pi^{+}$,
corresponding to (63.9$\pm$0.5)\% of all decays~\cite{PDG}, were utilized. 
The ionization energy loss $dE/dx$ in the TPC and the time of flight 
information of the particles from the time-of-flight detector~\cite{tof}
were used to select daughter pions and protons.
Cuts on decay topology, such as a distance of
the closest approach (DCA) between the trajectory of \lam(\alam)
candidates and the primary vertex, DCA between the two daughters, and
decay length of \lam(\alam) candidates were applied to reduce the
combinatoric background. Additional details about the \lam(\alam) reconstruction
can be found in Ref.~\cite{Adam:2018ivw}.

The longitudinal component of the polarization can be measured by projecting the
polarization onto the beam direction:
\begin{eqnarray}
P_z  &=& 
\frac{\langle\cos\theta_p^{\ast}\rangle}{\alpha_H
\langle\cos^2\theta_p^{\ast}\rangle}, \label{eq:pz1}
\end{eqnarray}
where $\theta_p^{\ast}$ is the polar angle of the daughter proton in
the \lam (\alam) rest frame and $\langle\rangle$ represents an average over
\lam (\alam) candidates in an event and then an average over all events.  
The decay parameter $\alpha_{H}$ is set to be 
$\alpha_{\Lambda}=-\alpha_{\bar{\Lambda}}=0.642\pm0.013$~\cite{PDG,note:alpha}.
%
%
If the detector has perfect acceptance and efficiency,  
$\langle\cos^2\theta_p^{\ast}\rangle$ leads to $1/3$.
In this study $\langle\cos^2\theta_p^{\ast}\rangle$ was extracted from
the data in order to account for pseudorapidity dependent detector acceptance effects.
This term was found to be close to $1/3$ for all centralities but showed a systematic decrease for lower track \pt. 
To extract the signal \costhe, two techniques were used: the event plane method and
the invariant mass method as described in Ref.~\cite{Adam:2018ivw}. In
the event plane method, \costhe was measured as a function of
azimuthal angle of $\lam(\alam)$ relative to $\Psi_2$. 
The average polarization along the beam direction is expected to be zero 
due to symmetry. Effects due to detector acceptance and inefficiencies 
are removed by subtracting the azimuthal average of \costhe 
from each azimuthal bin $i$ of \lam azimuthal angle: 
$\costhe^{\rm sub}_i=\costhe_i-\sum_i^{\rm n_{bin}}\costhe_i/{\rm n_{bin}}$.

\begin{figure}[hbt]
\begin{center}
\includegraphics[width=0.98\linewidth,clip]{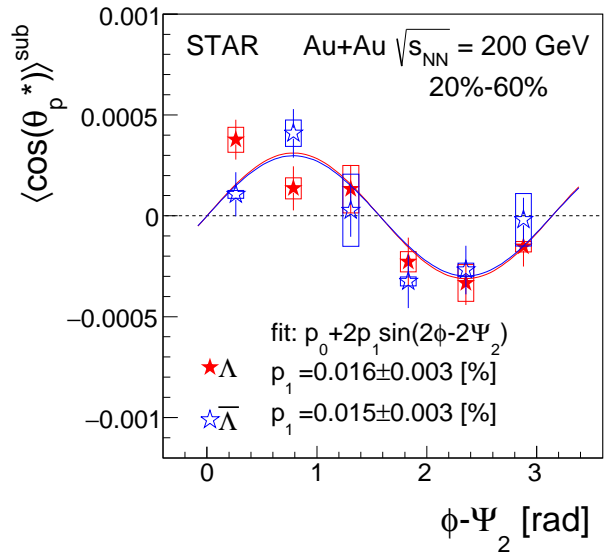}
\caption{\label{fig:cos}(Color online)
  $\langle\cos\theta_p^{\ast}\rangle$ of \lam and \alam hyperons as a
  function of azimuthal angle $\phi$ relative to the second-order
  event plane $\Psi_2$ for 20\%-60\% centrality bin in Au+Au
  collisions at \sqsn = 200 GeV. Open boxes show the systematic
  uncertainties and $\langle\rangle^{\rm sub}$ denotes the subtraction
  of the acceptance effect (see text).  Solid lines show the
  fit with the sine function shown inside the figure. 
  Note that the data are not corrected for the event plane resolution.}
\end{center}
\end{figure}

Figure~\ref{fig:cos} shows $\costhe^{\rm sub}$ of \lam and
\alam as a function of azimuthal angle relative to $\Psi_2$ for the
20\%--60\% centrality bin.  The solid lines indicate the fit results
to the function $p_0+2p_1\sin(2\phi-2\Psi_2)$, where $p_0$ and
$p_1$ are fit parameters.  The data are consistent with a sine
structure for both \lam and \alam as expected from the elliptic
flow. In the invariant mass method, the second-order Fourier sine
coefficient of $P_z$, $p_1=\langle P_z\sin(2\phi-2\Psi_2)\rangle$, was
measured as a function of the invariant mass. Following the same
procedure as described in Ref.~\cite{Adam:2018ivw}, the sine
coefficient was directly extracted. The extracted coefficient in both
methods was divided by Res($\Psi_2$) to account for the finite event
plane resolution. The invariant mass method was used to calculate the sine
coefficient of $P_z$ and the event plane method was used to cross-check and
provide an estimate of the systematic uncertainty.

\begin{figure}[hbt]
\begin{center}
\includegraphics[width=0.95\linewidth,clip]{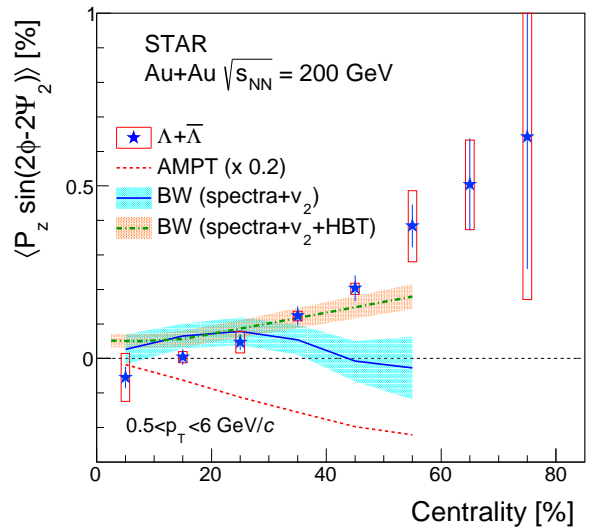}
\caption{\label{fig:pzct}(Color online) The second Fourier sine
  coefficient of the polarization of \lam and \alam along the beam
  direction as a function of the collision centrality in Au+Au
  collisions at \sqsn = 200 GeV.  Open boxes show the systematic
  uncertainties. Dotted line shows the AMPT calculation~\cite{Xia:2018tes}
  scaled by 0.2 (no \pt selection).  Solid and dot-dashed lines with the bands show the
  blast-wave (BW) model calculation for \pt~=~1 GeV/$c$ with \lam mass (see text for details).  }
\end{center}
\end{figure}

The systematic uncertainties were estimated by variation of the
topological cuts ($<2\%$), comparing the results from two methods for signal
extraction ($5\%$) as mentioned above, using different subevents 
($-1<\eta<-0.5$ and $0.5<\eta<1$) for $\Psi_2$ determination ($<11\%$), 
and	estimates of the possible background contribution to 
the signal ($4.3\%$). The numbers are for mid-central collisions.
%
%
Also the uncertainty from the decay parameter is accounted for (2\% for \lam and
9.6\% for \alam, see Ref.~\cite{Adam:2018ivw} for the detail).
We further studied the effect of a possible self-correlation between
the particles used for the \lam (\alam) reconstruction and the event
plane by explicitly removing the daughter particles from the event
plane calculation in Eq.~\eqref{eq:psi2}. There was no significant
difference between the results.  The \lam and \alam reconstruction 
efficiencies were estimated using {\sc GEANT}~\cite{Brun:1987ma} 
simulations of the STAR detector~\cite{tpc}. 
The correction is found to lower mean values of the $P_z$ sine coefficient by $\sim$10\%
in peripheral collisions and increases up to $\sim$50\% in central collisions, although
the variations are within statistical uncertainties.
No significant difference was observed between \lam and \alam as expected.  
Therefore, results from both samples were combined to reduce statistical uncertainties.

\begin{figure}[thb]
\begin{center}
\includegraphics[width=0.95\linewidth,clip]{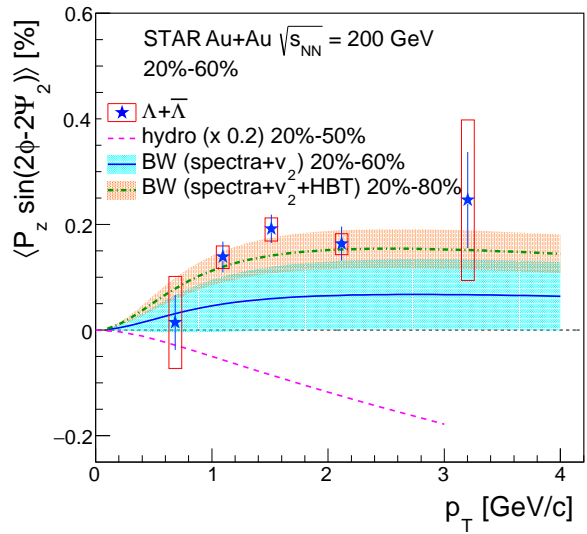}
\caption{\label{fig:pzpt}(Color online) The second Fourier sine
  coefficient of the longitudinal polarization of \lam and \alam
  hyperons as a function of \pt for 20\%-60\% centrality bin in Au+Au
  collisions at \sqsn = 200 GeV.  Open boxes show the systematic
  uncertainties.  Magenta dashed line shows the hydrodynamic model
  calculation~\cite{Becattini:2017gcx} scaled by 0.2.  Solid and
  dot-dashed lines with the bands show the blast-wave (BW) model
  calculations with \lam mass.}
\end{center}
\end{figure}

Figure~\ref{fig:pzct} presents the centrality dependence of the second
Fourier sine coefficient $\langle P_z\sin(2\phi-2\Psi_2)\rangle$.  
The increase of the signal with decreasing centrality is likely due to increasing
elliptic flow contributions in peripheral collisions. We note that, 
unlike elliptic flow, the polarization does disappear in the most central
collisions, where the elliptic flow is still significant due to initial 
density fluctuations.
Because of large uncertainties in
peripheral collisions, it is not clear whether the signal continues to
increase or levels off.  The results are compared to a
multiphase transport (AMPT) model~\cite{Xia:2018tes} as shown with
the dotted line.  The AMPT model predicts the opposite phase of the
modulations and overestimates the magnitude.
The blast-wave model study is discussed later.

Since the elliptic flow also depends on \pt as well as on the centrality,
the polarization may have \pt dependence. Figure~\ref{fig:pzpt} shows
the sine coefficients of $P_z$ as a function of the hyperon transverse
momentum.  No significant \pt dependence is observed for $\pt>1$
GeV/$c$, and the statistical precision of the single data point for $\pt<1$ GeV/$c$
is not enough to allow for definitive conclusions about the low \pt dependence.
In the hydrodynamic model calculation~\cite{Becattini:2017gcx}, 
the sine coefficient of $P_z$ increases in magnitude with \pt 
but shows the opposite sign to the data.

As shown in Figs.~\ref{fig:pzct} and \ref{fig:pzpt}, the hydrodynamic
and AMPT models predict the opposite sign in the sine coefficient
of the polarization and their magnitudes differ from the data roughly by a
factor of 5. The reason of this sign difference is under 
discussion in the community.  However, the sign change may be due to
the relation between azimuthal anisotropy and spatial anisotropy at
freeze-out~\cite{Voloshin:2017kqp}.  There could be contributions
from the kinematic vorticity originating from the elliptic flow as
well as from the temporal gradient of temperatures at the time of
hadronization~\cite{Becattini:2017gcx}.  A recent calculation using the
chiral kinetic approach predicts the same sign as the data~\cite{Sun:2018bjl}. 
The model accounts for the transverse component of the vorticity, resulting in axial charge currents. 
Note that both the hydrodynamic and transport models calculate local vorticity
at freeze-out and convert it to the polarization assuming local thermal
equilibrium of the spin degrees of freedom, while the chiral kinetic approach
takes into account nonequilibrium effects but does not consider a contribution from
the temperature gradient which is a main source of $P_z$ in the hydrodynamic model.

These models indicate that the contribution from the kinematic vorticity to $P_z$ is negligible
or opposite in the sign to the naive expectation from the elliptic flow.
In order to estimate the contribution from the kinematic vorticity we
employed the blast-wave model
(BW)~\cite{Schnedermann:1993ws,Adler:2001nb,Retiere:2003kf}.
%
Following Ref.~\cite{Retiere:2003kf} we parameterize the system velocity
field at freeze-out with temperature ($T$) and transverse flow rapidity ($\rho$) defined as
$\rho=\tilde{r}[\rho_0+\rho_2\cos(2\phi_b)]$. Here 
$\rho_0$ and $\rho_2$ are the maximal radial expansion rapidity and its azimuthal modulation,
$\tilde{r}$ is the relative distance to the edge of the source, and $\phi_b$ defines
the direction of the local velocity as indicated in Fig.~\ref{fig:cartoon}.  
The source shape, assumed to be elliptical in the transverse plane, is parameterized by 
the $R_y$ and $R_x$ radii. Boost invariance is assumed.  Two fits to the data are
performed: in one only spectra and elliptic flow of $\pi$, K, and p($\bar{p}$) are fit; the second
fit~\cite{Adams:2004yc} also includes azimuthal-angle-dependence of
the pion Gaussian source radii at freeze-out as measured via Hanbury-Brown and
Twiss (HBT) intensity interferometry.  The average longitudinal vorticity is calculated
according to the following formula:
%
%
\begin{eqnarray}
\mean{\omega_z\sin(2\phi)} &=&
\frac{\int d\phi_s \int rdr\, I_2(\alpha_t)K_1(\beta_t)\omega_z \sin(2\phi_b)} 
{\int d\phi_s \int rdr\, I_0(\alpha_t)K_1(\beta_t)} 
\\
\omega_z &=& \frac{1}{2} 
\left( \frac{\partial u_y}{\partial x} 
- \frac{\partial u_x}{\partial y}\right),
\end{eqnarray}
where the integration is over the transverse cross-sectional area of
the source, $u_\mu$ is a four-vector of the local flow
velocity~\cite{Retiere:2003kf}, $\phi_s$ is the azimuth of the
production point (see Fig.~\ref{fig:cartoon} for the relation to $\phi_b$), 
$\alpha_t=p_T/T\,\sinh \rho$, $\beta_t=m_T/T \cosh\rho$;
$I_n$ and $K_1$ are the modified Bessel functions.  Assuming a local
thermal equilibrium, the longitudinal component of the polarization is
estimated as $P_z \approx \omega_{z}/(2T)$. 
The uncertainties shown for the BW model calculations corresponds to
1 $\sigma$ variation in the model parameters.
See Ref.~\cite{BWstudy} for more details.

The BW calculations are compared to the data in Figs.~\ref{fig:pzct}
and \ref{fig:pzpt}.  From central to mid-central collisions both BW
calculations show positive sine coefficients which are compatible in both sign
and magnitude to the measurement, although the BW model
is based on a very simple picture of the freeze-out condition.  It was
shown in Ref.~\cite{Voloshin:2017kqp} that the vorticity in the BW
model has the effects of the velocity field anisotropy ($\rho_2/\rho_0$)
and the spacial source anisotropy ($R_y/R_x$) contributing with opposite
signs, which can explain a strong sensitivity of the BW model
predictions in the peripheral collisions to the inclusions of the HBT radii.

We have presented the first measurements of the 
longitudinal component of the polarization for \lam and \alam hyperons
in Au+Au collisions at \sqsn = 200 GeV. Finite signals of a quadrupole
modulation of both \lam and \alam polarization along the beam
direction are observed and found to be qualitatively consistent with the expectation
from the vorticity component along the beam direction due to the
elliptic flow.  The results exhibit a strong centrality dependence with
increasing magnitude as the collision centrality becomes more peripheral. 
No significant \pt
dependence is observed above $\pt>1$ GeV/$c$. A drop-off of the signal 
is hinted at for $\pt<1$ GeV/$c$.
The data were compared to calculations from hydrodynamic and
AMPT models, both of which show the opposite phase of the modulation
and overpredict the magnitude of the polarization. This might
indicate incomplete thermal equilibration of the spin degrees of
freedom for the beam direction component of the
vorticity/polarization, as it develops later in time compared to the
global polarization. On the other hand, the blast-wave model
calculations are much closer to the data, even more so when 
the azimuthally sensitive HBT results along with the \pt spectra and $v_2$ 
are included in the model fit.
The blast-wave model predicts the correct
phase of $P_z$ modulation and a similar \pt dependence; the version
with HBT radii included in the fit also reasonably describes the
centrality dependence.  These results together with the results of the global polarization 
may provide information on the relaxation time needed to convert the vorticity to particle polarization.
Further theoretical and experimental studies are needed for better understanding.  
%
%
%
%
%
\begin{acknowledgments}
We thank the RHIC Operations Group and RCF at BNL, the NERSC Center at
LBNL, and the Open Science Grid consortium for providing resources and
support. This work was supported in part by the Office of Nuclear
Physics within the U.S. DOE Office of Science, the U.S. National
Science Foundation, the Ministry of Education and Science of the
Russian Federation, National Natural Science Foundation of China,
Chinese Academy of Science, the Ministry of Science and Technology of
China and the Chinese Ministry of Education, the National Research
Foundation of Korea, Czech Science Foundation and Ministry of
Education, Youth and Sports of the Czech Republic, 
Hungarian National Research, Development and Innovation Office (FK-123824), 
New National Excellency Programme of the Hungarian Ministry of Human Capacities (UNKP-18-4), 
Department of Atomic Energy and Department of Science and Technology of the Government of India, 
the National Science Centre of Poland, the Ministry of Science, Education and Sports of the Republic of Croatia, 
RosAtom of Russia and German Bundesministerium fur Bildung, Wissenschaft, Forschung and Technologie (BMBF) and the Helmholtz Association.
\end{acknowledgments}
%
\bibliography{ref_zpol}   
\end{document}

%% file: authorlist_05282019.tex
\affiliation{Abilene Christian University, Abilene, Texas   79699}
\affiliation{AGH University of Science and Technology, FPACS, Cracow 30-059, Poland}
\affiliation{Alikhanov Institute for Theoretical and Experimental Physics, Moscow 117218, Russia}
\affiliation{Argonne National Laboratory, Argonne, Illinois 60439}
\affiliation{American Univerisity of Cairo, Cairo, Egypt}
\affiliation{Brookhaven National Laboratory, Upton, New York 11973}
\affiliation{University of California, Berkeley, California 94720}
\affiliation{University of California, Davis, California 95616}
\affiliation{University of California, Los Angeles, California 90095}
\affiliation{University of California, Riverside, California 92521}
\affiliation{Central China Normal University, Wuhan, Hubei 430079 }
\affiliation{University of Illinois at Chicago, Chicago, Illinois 60607}
\affiliation{Creighton University, Omaha, Nebraska 68178}
\affiliation{Czech Technical University in Prague, FNSPE, Prague 115 19, Czech Republic}
\affiliation{Technische Universit\"at Darmstadt, Darmstadt 64289, Germany}
\affiliation{E\"otv\"os Lor\'and University, Budapest, Hungary H-1117}
\affiliation{Frankfurt Institute for Advanced Studies FIAS, Frankfurt 60438, Germany}
\affiliation{Fudan University, Shanghai, 200433 }
\affiliation{University of Heidelberg, Heidelberg 69120, Germany }
\affiliation{University of Houston, Houston, Texas 77204}
\affiliation{Huzhou University, Huzhou, Zhejiang  313000}
\affiliation{Indian Institute of Science Education and Research, Tirupati 517507, India}
\affiliation{Indiana University, Bloomington, Indiana 47408}
\affiliation{Institute of Physics, Bhubaneswar 751005, India}
\affiliation{University of Jammu, Jammu 180001, India}
\affiliation{Joint Institute for Nuclear Research, Dubna 141 980, Russia}
\affiliation{Kent State University, Kent, Ohio 44242}
\affiliation{University of Kentucky, Lexington, Kentucky 40506-0055}
\affiliation{Lawrence Berkeley National Laboratory, Berkeley, California 94720}
\affiliation{Lehigh University, Bethlehem, Pennsylvania 18015}
\affiliation{Max-Planck-Institut f\"ur Physik, Munich 80805, Germany}
\affiliation{Michigan State University, East Lansing, Michigan 48824}
\affiliation{National Research Nuclear University MEPhI, Moscow 115409, Russia}
\affiliation{National Institute of Science Education and Research, HBNI, Jatni 752050, India}
\affiliation{National Cheng Kung University, Tainan 70101 }
\affiliation{Nuclear Physics Institute of the CAS, Rez 250 68, Czech Republic}
\affiliation{Ohio State University, Columbus, Ohio 43210}
\affiliation{Institute of Nuclear Physics PAN, Cracow 31-342, Poland}
\affiliation{Panjab University, Chandigarh 160014, India}
\affiliation{Pennsylvania State University, University Park, Pennsylvania 16802}
\affiliation{NRC "Kurchatov Institute", Institute of High Energy Physics, Protvino 142281, Russia}
\affiliation{Purdue University, West Lafayette, Indiana 47907}
\affiliation{Pusan National University, Pusan 46241, Korea}
\affiliation{Rice University, Houston, Texas 77251}
\affiliation{Rutgers University, Piscataway, New Jersey 08854}
\affiliation{Universidade de S\~ao Paulo, S\~ao Paulo, Brazil 05314-970}
\affiliation{University of Science and Technology of China, Hefei, Anhui 230026}
\affiliation{Shandong University, Qingdao, Shandong 266237}
\affiliation{Shanghai Institute of Applied Physics, Chinese Academy of Sciences, Shanghai 201800}
\affiliation{Southern Connecticut State University, New Haven, Connecticut 06515}
\affiliation{State University of New York, Stony Brook, New York 11794}
\affiliation{Temple University, Philadelphia, Pennsylvania 19122}
\affiliation{Texas A\&M University, College Station, Texas 77843}
\affiliation{University of Texas, Austin, Texas 78712}
\affiliation{Tsinghua University, Beijing 100084}
\affiliation{University of Tsukuba, Tsukuba, Ibaraki 305-8571, Japan}
\affiliation{United States Naval Academy, Annapolis, Maryland 21402}
\affiliation{Valparaiso University, Valparaiso, Indiana 46383}
\affiliation{Variable Energy Cyclotron Centre, Kolkata 700064, India}
\affiliation{Warsaw University of Technology, Warsaw 00-661, Poland}
\affiliation{Wayne State University, Detroit, Michigan 48201}
\affiliation{Yale University, New Haven, Connecticut 06520}

\author{J.~Adam}\affiliation{Creighton University, Omaha, Nebraska 68178}
\author{L.~Adamczyk}\affiliation{AGH University of Science and Technology, FPACS, Cracow 30-059, Poland}
\author{J.~R.~Adams}\affiliation{Ohio State University, Columbus, Ohio 43210}
\author{J.~K.~Adkins}\affiliation{University of Kentucky, Lexington, Kentucky 40506-0055}
\author{G.~Agakishiev}\affiliation{Joint Institute for Nuclear Research, Dubna 141 980, Russia}
\author{M.~M.~Aggarwal}\affiliation{Panjab University, Chandigarh 160014, India}
\author{Z.~Ahammed}\affiliation{Variable Energy Cyclotron Centre, Kolkata 700064, India}
\author{I.~Alekseev}\affiliation{Alikhanov Institute for Theoretical and Experimental Physics, Moscow 117218, Russia}\affiliation{National Research Nuclear University MEPhI, Moscow 115409, Russia}
\author{D.~M.~Anderson}\affiliation{Texas A\&M University, College Station, Texas 77843}
\author{R.~Aoyama}\affiliation{University of Tsukuba, Tsukuba, Ibaraki 305-8571, Japan}
% ...

\collaboration{STAR Collaboration}\noaffiliation